\documentclass{ws-ijbc}
\usepackage{ws-rotating}     % used only when sideways tables/figures are used
\usepackage{graphicx}
\usepackage{epstopdf}
\usepackage{xcolor, soul}
\usepackage{dcolumn}% Align table columns on decimal point
\usepackage{bm}% bold math
\usepackage{mathrsfs} % script-like, curvy letters.
\usepackage{amsmath} %curvy letters.

\begin{document}

\catchline{}{}{}{}{} % Publisher's Area please ignore

\markboth{A. Ray and D. Ghosh}{Another new chaotic system: bifurcation and chaos control}

\title{Another new chaotic system: bifurcation and chaos control}

\author{Arnob Ray and Dibakar Ghosh\footnote{dibakar@isical.ac.in}}

\address{Physics and Applied Mathematics Unit, Indian Statistical Institute, Kolkata 700108, India}
\maketitle

\begin{abstract}
     We propose a new simple three-dimensional continuous autonomous model with two nonlinear terms and observe the dynamical behavior with respect to system parameter. This system changes the stability of fixed point via Hopf bifurcation and then undergoes a cascade of period-doubling route to chaos. We analytically derive the first Lyapunov coefficient to investigate the nature of Hopf bifurcation and also investigate well-separated regions for different kind of attractors in two-dimensional parameter space. Next, we introduce a time{-}scale ratio parameter and calculate the slow manifold using geometric singular perturbation theory. Finally, the chaotic state is annihilated by decreasing the value of time{-}scale ratio parameter.   
\end{abstract}

\keywords{Chaos, Hopf-bifurcation, first Lyapunov coefficient, slow-fast dynamics.}

%\begin{multicols}{2}
	
\section{Introduction}
  Finding a new chaotic system began from the last six decades to understand different natural and man made systems. Still scientific communities are investigating new chaotic system due to its practical applications in engineering \cite{szemplinska1991}, finance \cite{guegan2009}, plasma \cite{ray2008, ray2009}, time series analysis \cite{kodba2005, nazarimehr2018a} and natural observations like, climatology \cite{selvam2007}, biology \cite{may1987, ghosh2007}, geology \cite{turcotte1997} etc. After discovery of Lorenz model \cite{lorenz1963} in 1963, there are several chaotic models were investigated like,  R\"{o}ssler system \cite{rossler1976}, Chen system \cite{chen1999}, Chua's attractor \cite{chua1986}, Sprott system \cite{sprott1994, sprott2000}, logistic model \cite{may1976}, predator-prey based model \cite{hastings1991, ginoux2019} and many more. At first, theoretical background behind generation of chaotic solution in deterministic system was not developed. But, now different ways are brought to light to understand and realize the chaotic attractors and its dynamics, both from analytical as well as numerical approaches \cite{eckmann1985, perc2005, silva2018}. Initially, three well-known routes were reported for generating chaos in dynamical systems, $\it i.e.$, period-doubling, quasi-periodic and intermittent routes \cite{ott2002}. Nowadays, researchers are  fascinated for finding chaotic attractor in a different manner.  One interesting topic is to find multistable chaotic attractors \cite{bayani2019} which has important application and impact to explore a broad field of research \cite{dudkowski2016}. Besides a new concept of attractor arises connected with multistability like attractors with no equilibria \cite{wei2011, jafari2013b, pham2014}, line equilibria \cite{jafari2013a, nazarimehr2018b}, plane of equilibrium \cite{jafari2016} or coexistence of more than one chaotic attractors \cite{dudkowski2016, ray2009} etc. Basin of attraction of chaotic attractor does not touch with small neighborhood of the unstable fixed point, which is quite different from the traditional chaotic attractors (called {\it self-excited chaotic attractors}). This kind of attractor is familiar as {\it hidden attractor} as finding of hidden attractor is difficult task usually \cite{dudkowski2016}. Researches are also eager to control of the multistability \cite{pisarchik2014}. But finding a simple chaotic system is still a challenging task with a great impact for basic research \cite{liu2004, lu2002, wang2012}. We have also knowledge about chaotic systems in which different timescales are separated which have huge impact in various fields,  like chemical oscillatory model \cite{epstein1983}, Hindmarsh-Rose neuronal model \cite{hindmarsh1984}, prey-predator model \cite {kuwamura2009} and so on. Well known features of chaotic attractor are sensitive dependence of initial conditions, boundedness, aperiodic and appearance of it is confirmed by existence of positive Lyapunov exponent in the parameter region \cite{strogatz2014}.  Researchers are also very much interested to study the chaotic dynamics in delay systems due to its impactful applicability \cite{wei2019, khajanchi2018, banerjee2019, ghosh2017}. With studying about generation of chaotic system,  parallel researchers have been focused about control of chaos in non-delay chaotic systems \cite{ott1990, perc2006, jun2008} as well as delay chaotic systems \cite{ghosh2008, bhowmick2014}. Its application includes chemical reaction oscillations, turbulent fluids etc. \cite{boccaletti2000}. From the above discussion, the finding of a new chaotic system and exploring different dynamical states deserve a special attention.

\par In this paper, we propose a simple nonlinear system which exhibits single-scroll chaotic attractor in a certain parameter region. At first, we calculate the fixed points of the system and analyze their linear stability analyses. From these analyses, we observe that the two fixed points out of three always unstable, whereas one fixed point changes its stability via Hopf bifurcation and finally goes to chaotic state through period-doubling route. We also analytically derive the first Lyapunov coefficient to check the stability of the limit cycle. Then, we introduce slow-fast ratio parameter in the proposed model and investigate the previously mentioned system parameter region by varying this slow-fast parameter. We analytically calculate the slow manifold with the help of geometric singular perturbation theory. The chaotic behavior completely annihilate if the time scale ratio parameter is low.
\par The rest of the paper is organized as follows: Section-2 devotes the introduction of a new chaotic system and linear stability analysis of the fixed points. We also discuss the bifurcation diagram, Lyapunov exponent and two-dimensional parameter regions by varying the system parameters. In Sec-3, we discuss the effect of slow-fast parameter and controlling of chaos. Finally, conclusions are drawn in Sec-4.     

\section{Model and system dynamics}
We propose a three-dimensional model with two nonlinear turns in the following form,
\begin{equation}
\begin{array}{l}\label{eq.1}	
\dot{x} =-x+y+z,\\
\dot{y} =xy-z,\\
\dot{z} =-axz+y+b,\\
\end{array}
\end{equation}
where $a,~b$ are two positive parameters and $x, y$ and $z$ are state variables. Here dot denotes time derivative with respect to time $t$. We first study the linear stability analysis of the equilibrium points of the above system. The equilibrium points are in the form $(x^*, f(x^*), g(x^*))$, where $f(x)$ and $g(x)$ are given by
\[
f(x)=\frac{x}{1+x},~~~g(x)=\frac{x^2}{1+x}.
\]
Here $x^*$ satisfies the following equation
\begin{equation}
\begin{array}{l}\label{eq.2}	
ax^*{^3}-(b+1)x^*-b=0.
\end{array}
\end{equation}
Solving Eqn. (\ref{eq.2}), we have three values of $x^*$ corresponding to three roots $x_1^*$, $x_2^*$ and $x_3^*$,  
\begin{equation*}
\begin{array}{l}
x_1^*=2\sqrt{\frac{b+1}{3a}}cos(\frac{\theta}{3}),\\
x_2^*=-2\sqrt{\frac{b+1}{3a}}cos(\frac{\pi-\theta}{3}),\\
x_3^*=-2\sqrt{\frac{b+1}{3a}}cos(\frac{\pi+\theta}{3}),
\end{array}
\end{equation*}
\begin{figure}[ht]
	\centerline{
		\includegraphics[scale=0.6]{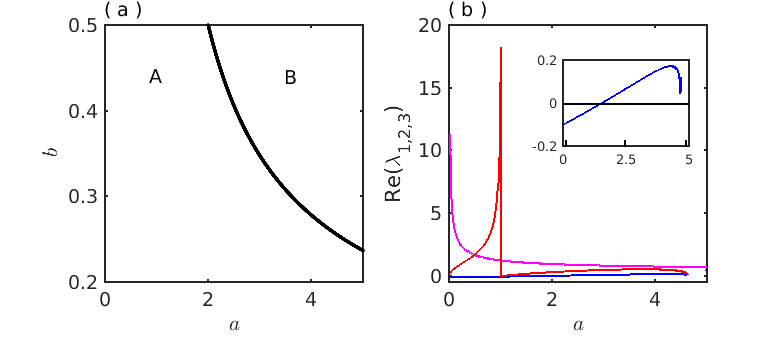} 	}
	\caption{(a) $a$-$b$ parameter space where region A denotes the existence of three fixed points, whereas one fixed point exists in the region B. (b) Variation of maximum real part of the eigenvalues corresponding  to the three fix points by varying the parameter $a$. Here maximum eigenvalues corresponding to two points $x_1^*$ and $x_2^*$ are always positive (red and magenta curves) and $x_3^*$ is negative up to $a<1.53$ and change the sign after that which is shown in inset figure. Here one fixed point changes its stability and other two are always unstable. Here $b=0.25$.} 
	\label{fig1} 
\end{figure}
where
\[
\theta=\tan^{-1}\Big(\frac{1}{3b}\sqrt{\dfrac{4(b+1)^3-27ab^2}{3a}}\Big).
\]
The above three roots of the Eqn. (\ref{eq.2}) are real if  
\[
0<a<\frac{4(b+1)^3}{27b^2}.
\] 
  
In Fig.\ \ref{fig1}(a), we draw a curve $a=\frac{4(b+1)^3}{27b^2}$. Left side of the curve in $a$-$b$ parameter space (denoted by A), system possesses three fixed points and only one fixed point exists at the right side of the curve (denoted by B). Now, we check the qualitative nature of equilibrium point $(x^*, f(x^*), g(x^*))$ of the system (\ref{eq.1}) by linear stability analysis. The Jacobian matrix $J$ of the system at the equilibrium point $(x^*, f(x^*), g(x^*))$ is as follows,
\[
J=
\left[ {\begin{array}{ccc}
	-1   &   1   &   1 \\
	f(x^*)   &   x^*   &   -1 \\
	-ag(x^*)   &   1   &   -ax^*\\
	\end{array} } \right].
\]	

So, the characteristic equation of the matrix $J$ is\\

\begin{equation}
\begin{array}{l}\label{eq.3}
\lambda^3+a_2\lambda^2+a_1\lambda+a_0=0,
\end{array}
\end{equation}
where\\
\begin{equation*}
	\begin{array}{l}
		a_2=1-x^*+ax^*,\\
		a_1=\dfrac{1-ax^*{^3}+(a-1)x^*{^2}+(a-1)x^*}{1+x^*},\\
		a_0=\dfrac{1-3ax^*{^2}-2ax^*{^3}}{1+x^*}.
	\end{array}
\end{equation*}
Solving Eqn. (\ref{eq.3}), the three eigenvalues are
\begin{equation}
\begin{array}{l}\label{eq.4}
\lambda_1=(z_1+z_2)-\frac{a_2}{3},\\
\lambda_2=-\frac{z_1+z_2}{2}-\frac{a_2}{3}+i\frac{\sqrt3(z_1-z_2)}{2},\\
\lambda_3=-\frac{z_1+z_2}{2}-\frac{a_2}{3}-i\frac{\sqrt3(z_1-z_2)}{2},\\
\end{array}
\end{equation}
where
$
z_1=(q+(p^3+q^2)^\frac{1}{2})^\frac{1}{2}$, $ z_2=(q-(p^3+q^2)^\frac{1}{2})^\frac{1}{2}$, $p=\frac{a_1}{3}-\frac{a_2^2}{9}$ and $ q=\frac{a_1a_2-3a_0}{6}-\frac{a_2^3}{27}$.
If $p^3+q^2>0$, Eqn.\ (\ref{eq.3}) gives one real root with a pair of complex conjugate eigenvalues and another case, $i.e.$,  $p^3+q^2<0$, the equation gives three  real roots.

\par Now, we fix the system parameter $b$ at $b=0.25$ and consider $a$ as a bifurcation parameter. In Fig.\ \ref{fig1}(b), we plot the  maximum real part of the eigenvalues of $J$ at $x_1^*, x_2^*,$ and $x_3^*$ with respect to the system parameter $a$. From this figure, the maximum real part of the eigenvalue (Re(${\lambda_{3}}$)) of $J$ at third root, $i.e.,~x_3^*$ shows negative in the range $a \in$ (0, 1.53) (blue line) and after that turns to positive (clearly shown in the inset figure of Fig.\ \ref{fig1}(b)), but for remaining two roots, $i.e.,~x_1^*,~x_2^*$, their associated maximum eigenvalues (Re(${\lambda_{1,2}}$)) are always positive (magenta and red lines). Interestingly, we see that at $a=4.63$, there are one maximum real part of the eigenvalues corresponding to $x_1^*, x_2^*,$ and $x_3^*$ exists. This means for $a>4.63$, the Eqn. (\ref{eq.2}) possesses only one real root ($B$ region in Fig.\ \ref{fig1}(a)). So it can be concluded that the Eqn. (\ref{eq.1}) has only one stable fixed point corresponding to $x_3^*$ in the range $a \in (0, 1.53)$. Now, for a specific value of $a=0.5$ and $b=0.25$, we calculate $x^*$ and associated eigenvalues  $\lambda_{1, 2, 3}$ as follows:
\begin{equation*}
	\begin{array}{l}
		x_1^*: x^*\approx1.6731,\\ (\lambda_{1},\lambda_{2},\lambda_{3})\approx(1.6791,-0.9213\pm0.9524i);\\
		x_2^*: x^*\approx-1.4698,\\ (\lambda_{1},\lambda_{2},\lambda_{3})\approx(-3.0764,1.718,-0.3765);\\  
		x_3^*: x^*\approx-0.2034,\\  (\lambda_{1},\lambda_{2},\lambda_{3})\approx(-0.9653,-0.0682\pm1.1072i).      
	\end{array}
\end{equation*}

\begin{figure*}[ht]
	\centerline{
		\includegraphics[scale=0.31]{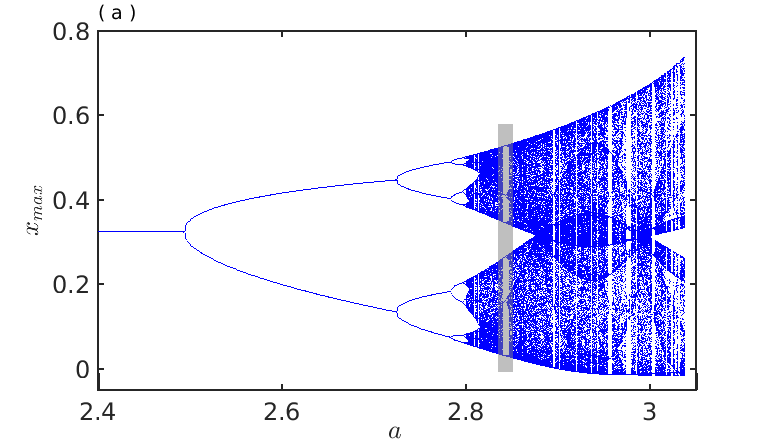}
		\includegraphics[scale=0.31]{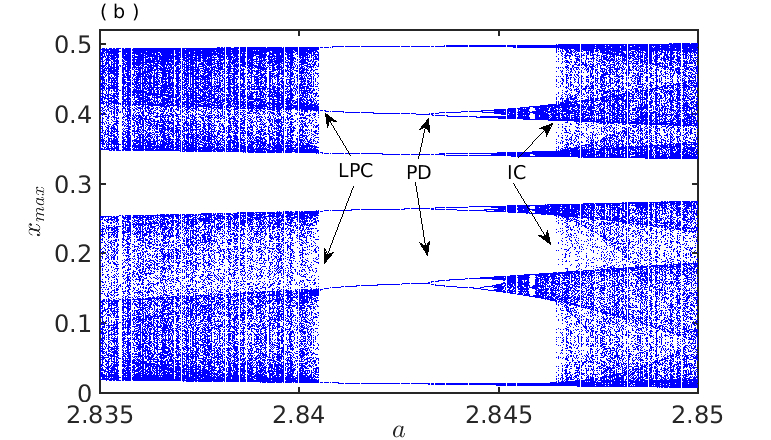}
		\includegraphics[scale=0.31]{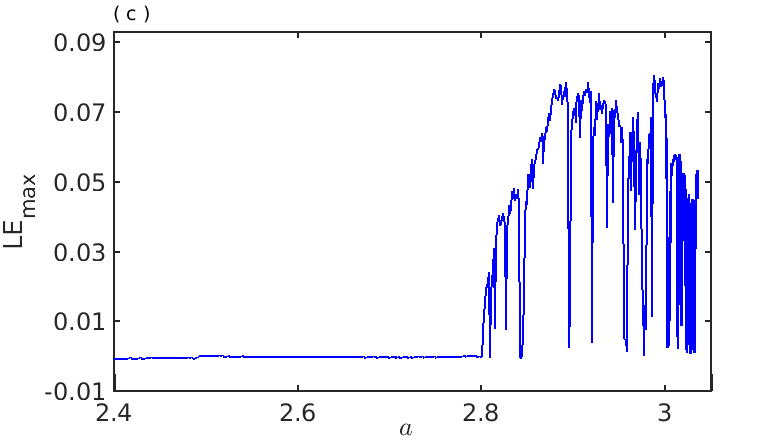}}
	\caption{(a) Bifurcation diagram of the new system Eqn.\ (\ref{eq.1}) for $a\in[2.4,3.037]$ showing period-doubling route to chaos.
		(b)	Zoom version of a periodic window shown as shaded region in figure (a). Here LPC, PD, and IC denote the limit point bifurcation of cycles, period-doubling, and interior crisis, respectively. (c) Corresponding maximum Lyapunov exponent with respect to $a\in[2.4,3.037]$ to verify different states (periodic state and chaotic state). Here, we fix $b=0.25$. Initial condition is taken as $(x(0), y(0), z(0))=(0.3, 0.2, 0.1).$}
	\label{fig2} 
\end{figure*}
\begin{figure}[ht]
	\centerline{
		\includegraphics[scale=0.55]{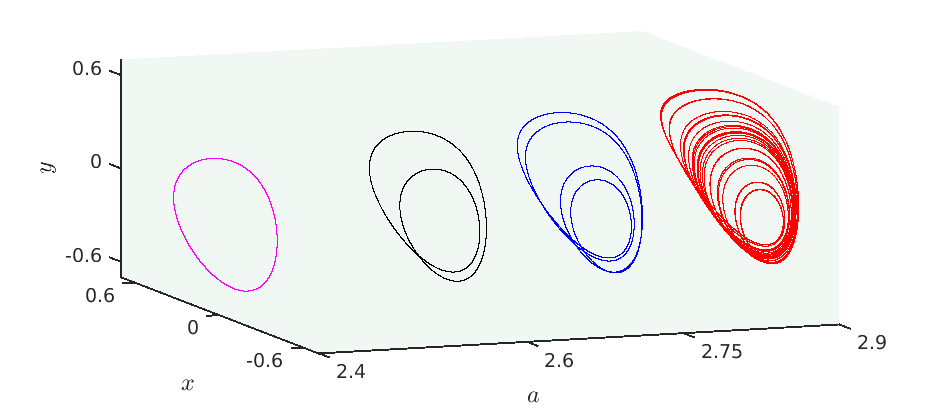}}
	\caption{Phase portraits for different values of $a$: $a=2.4$, period-1 (magenta); $a=2.6$, period-2 (black); $a=2.75$, period-4 (blue) and $a=2.9$, chaotic state (red). From the phase diagrams, period-doubling route to chaos is easily observed.  Here $b=0.25$. }
	\label{fig3} 
\end{figure}
Using the sign of eigenvalues, the fixed points corresponding to $x_1^*, x_2^*,$ and $x_3^*$ of the system (\ref{eq.1}) are unstable saddle focus, unstable saddle and stable focus, respectively at $a=0.5$ and $b=0.25$. So we focus on the fixed point corresponding to $x_3^*$ and study Hopf bifurcation analysis. The characteristic equation at the point $x_3^*$ becomes,
\begin{equation}
\begin{array}{l}\label{eq.5}
\lambda^3+r_2\lambda^2+r_1\lambda+r_0=0,
\end{array}
\end{equation}
where
\begin{equation*}
	\begin{array}{l}
		r_2=1+(1-a)(\frac{\sqrt{15a}}{3a})cos(\theta_1),\\
		r_1=\dfrac{1-\frac{(a-1)\sqrt{15a}}{3a}cos(\theta_1)+\frac{5(a-1)}{3a}cos^2\theta_1+\frac{5\sqrt{15a}}{9a}cos^3\theta_1}{1-(\frac{\sqrt{15a}}{3a})cos(\theta_1)},\\
		r_0=\dfrac{1-5cos^2(\theta_1)+\frac{10\sqrt{15a}}{9a}cos^3(\theta_1)}{1-(\frac{\sqrt{15a}}{3a})cos(\theta_1)},
	\end{array}
\end{equation*}
and,
\[
\theta_1=\frac{\sqrt{3a(125-27a)}}{9a}.
\]
According to the Routh-Hurwitz stability criterion, the real parts of all the eigenvalues $\lambda$ become negative if %Finally, $l_1=0.04196~(> 0)$. \\
$r_2>0, r_0 >0,$ and $r_1r_2 >r_0$. Numerically, we calculate the Hopf bifurcation point at $a^* \approx 1.524053$ where the real parts of the complex conjugate roots of Eqn. (\ref{eq.5}) becomes zero \cite{kuznetsov2004}.
% From similar calculation as (\ref{eq.4}), condition of geting the critical point of Hopf bifurcation parameter $a$ as follows,
%\begin{equation*}
%\begin{array}{l}
%-\frac{1}{2}[(q_1+(p_1^3+q_1^2)^\frac{1}{2})^\frac{1}{2}+(q_1-(p_1^3+q_1^2)^\frac{1}{2})^\frac{1}{2}]-\frac{r_2}{3}=0
%\end{array}
%\end{equation*}
% with $p_1^3+q_1^2>0$. Here,
% \[
% p_1=\frac{r_1}{3}-\frac{r_2^2}{9}, ~~~~ q_1=\frac{r_1r_2-3r_0}{6}-\frac{r_2^3}{27}.\\
% \]
%Numerically, we get the Hopf bifurcation point $a^* \approx 1.524053$ and at that point, roots of equation (\ref{eq.5}) are
%\begin{equation*}
%\begin{array}{l}
%(\lambda_{1},\lambda_{2},\lambda_{3})\approx(-0.8891,\pm1.08431i).      
%\end{array}
%\end{equation*}
\begin{figure}[ht]
	\centerline{
		\includegraphics[scale=0.65]{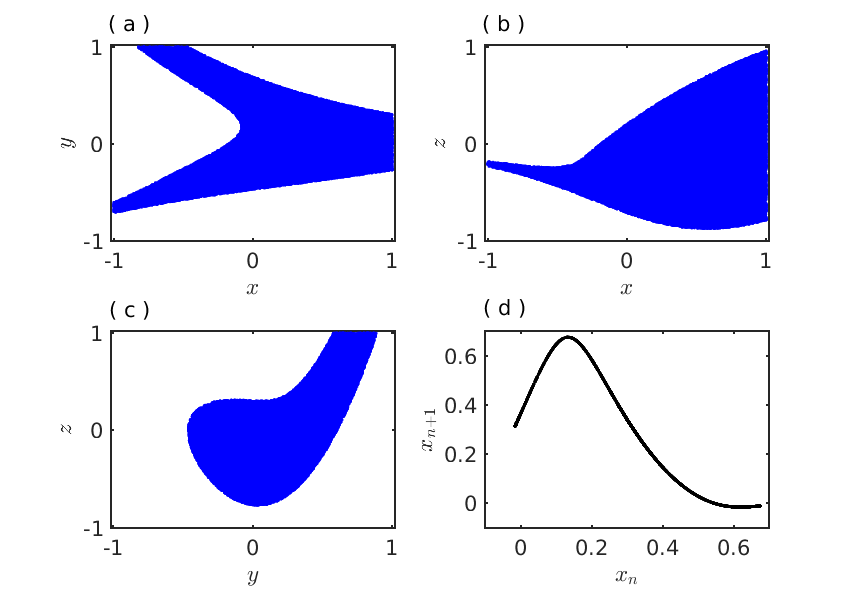} 	}
	\caption{ Basin of attraction of chaotic attractor at (a) $z=0.1$ plane, (b) $y=0.1$ plane, and (c) $x=0.1$ plane. White region represents the unbounded state. (d) First return map for the chaotic attractor. This map shows unimodal behavior which signify period-doubling route to chaos. Initial condition is chosen at $x(0)=0.3, y(0)=0.2,$ and $z(0)=0.1$. Parameter values are fixed at $a=2.9$ and $b=0.25$}
	\label{fig4} 
\end{figure} 

Now, we discuss about the nature of Hopf bifurcation \cite{kuznetsov2004}. For this purpose, first we check the transversality condition and then calculate the value of first Lyapunov coefficient \cite{kuznetsov2004, mello2009, wei2014}. The transversality condition, $\ i.e.$, real part of $\frac{d\lambda(a^*)}{da}|_{\lambda=1.0843i} \approx 0.06839 (\ne 0)$ is satisfied here. The first Lyapunov coefficient is $l_1\approx0.04196~(> 0)$ (detail calculations are in appendix). So the fixed point (corresponding to $x_3^*$) deals with subcritical Hopf bifurcation at $a^* \approx 1.524053$ and it indicates that there is an unstable limit cycle exists in the neighborhood of the fixed point.

\par  Next, we study the dynamical behavior of the equilibrium point corresponding to $x_3^*$ with respect to the parameter $a$ after crossing the value $a=a^*$. After crossing $a^*$, the fixed point loses its stability and a stable limit cycle of period-1 appears and persists up to $a\approx2.493$. Then we achieve chaotic dynamics via period-doubling (PD) route \cite{strogatz2014}. This scenario is depicted in Fig.\ \ref{fig2}(a). At $a\approx3.037$, this chaotic attractor is destructed due to boundary crisis \cite{grebogi1983} as the chaotic attractor collides with unstable fixed point (corresponding to $x_1^*$). We take a zoom version of periodic window  at the neighborhood of $a=2.845$ (shaded region in Fig.\ \ref{fig2}(a)) in Fig.\ \ref{fig2}(b). System changes from chaotic oscillation to periodic oscillation through limit point bifurcation of cycles (LPC). Chaotic oscillation again revives via PD bifurcation and amplifies via interior crisis (IC). Then we plot maximum Lyapunov exponent ($\rm LE_{max}$) in Fig.\ \ref{fig2}(c) for verifying our result in Fig.\ \ref{fig2}(a). Positive value of $\rm LE_{max}$ after a certain value of $a$ signifies the presence of chaos. For four different values of $a$, (i.e., $a=2.4, 2.6, 2.75$ and $2.9$), we display four different kind of attractors, $\ i.e.$, limit cycle of period-1 (magenta), period-2 (black), period-4 (blue), and chaos (red), respectively. The occurrence of period-doubling cascade for generating chaos is verified by calculating Feigenbaum universal constant \cite{strogatz2014}. The ratio of distances between successive intervals between the bifurcate values of the system parameter, represented by a constant factor
\[
\delta=\lim_{k\to\infty} \frac{a_{k+1}-a_{k}}{a_{k+2}-a_{k+1}}
\] 
approaches to 4.66992.... The value of Feigenbaum constant for our case are given in the following table: 
\vspace{0.1cm}
\begin{table}[h!]
	\centering
	\begin{tabular}{||c c c c||} 
		\hline
		k & Period & Bifurcation parameter $a_k$ & Ratio $\frac{a_{k+1}-a_{k}}{a_{k+2}-a_{k+1}}$ \\ [1ex] 
		\hline\hline
		1 & $2^1$ & 2.491 & Not applicable \\ 
		2 & $2^2$ & 2.7234 & Not applicable \\
		3 & $2^3$ & 2.7822 & 3.952 \\
		4 & $2^4$ & 2.7966 & 4.0833 \\
		5 & $2^5$ & 2.79969 & 4.6602 \\
		6 & $2^6$ & 2.8003517 & 4.6696 \\ [1ex] 
		\hline
	\end{tabular}
\end{table}\vspace{1.0cm}
\begin{figure}[ht]
	\centerline{
		\includegraphics[scale=0.45]{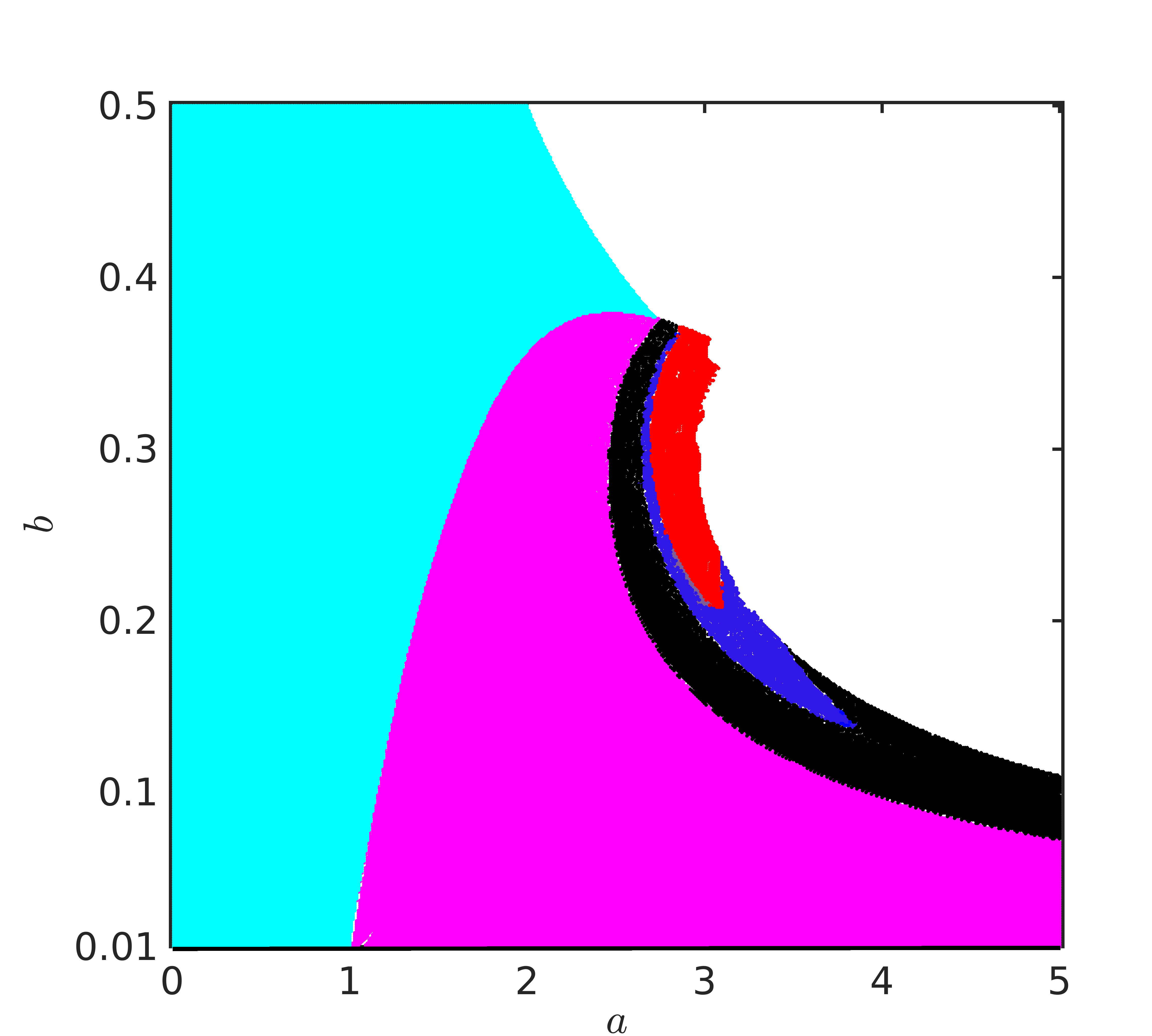} 	}
	\caption{ Different dynamical behaviors in the $a$-$b$ parameter space. Here Cyan:  stable steady state, magenta: period-1, black: period-2, blue: period-4, and red: greater than period-4 or chaotic state. White region is for unbounded solution. The initial condition is $x(0)=0.3, y(0)=0.2$, and $z(0)=0.1$.}
	\label{fig5} 
\end{figure}

 Next, we identify the region of initial conditions for the chaotic state. The basin of attractions at $z=0.1$ plane, $y=0.1$ plane, and $x=0.1$ plane  for the chaotic attractor at $(a, b)=(2.9, 0.25)$ are depicted in Figs.\ \ref{fig4}(a-c). We get fully connected basin of attractions for respective three cases.  At the same set of parameter value, we also plot first return map, $\ i.e.$, $x_n$ vs. $x_{n+1}$ in Fig.\ \ref{fig4}(d), where $x_{n}$ denotes the $n$-th local maxima of the time series. Unimodal nature of this first return map indicates the PD bifurcation route to generate chaos. 
\par To explore the complete study by varying the two parameters $a$ and $b$ simultaneously, we plot a two parameter $(a$-$b)$ bifurcation diagram \cite{ghosh2015} for the range $a \in [0,5]$ and $b \in [0.01,0.5]$ in Fig.\ \ref{fig5}. This demonstrates the whole dynamics of the system (\ref{eq.1}). In this phase diagram, cyan, magenta, black, blue, and red regions represent steady state, limit cycle of period-1, period-2, period-4, and greater than period-4 (or chaotic),  respectively. White portion represents unbounded solution for higher values of $a$ and $b$. Interestingly, well-separated parameter space is observed in $(a$-$b)$ bifurcation diagram. 
   
\section{Slow-fast dynamical system and controlling chaos}
 In this section, we introduce a slow-fast time scale ratio parameter to create slow-fast dynamics in the proposed system. We try to understand the effect of this time scale parameter. A general framework of a slow-fast dynamical system \cite{guckenheimer2008} can be written as,
	\begin{equation*}
	\begin{array}{lll}
				
	\epsilon \dfrac{dx}{dt}=f(x,y,\eta); 
	~\dfrac{dy}{dt}= g(x,y,\eta),
	\end{array}
	\end{equation*}
	where $x\in \mathbb{R}^{m} \;\mbox{and}\; y\in \mathbb{R}^{n}$ are fast and slow variables, respectively, $\eta\in \mathbb{R}^{p}$ is a model parameter, $0<\epsilon<1$ represents the ratio of time-scales. 	Setting $\epsilon$=0, the trajectory of above equation converges to the solution of differential algebraic equation $f(x,y,\eta) =0$ and $\dot {y}=g(x,y,\eta)$, where 	$S_{0}= \{(x,y) \in \mathbb{R}^{m} \times \mathbb{R}^{n}| f(x,y,\eta) =0\}$ is a critical manifold.

\par Now, in our model we introduce a time-scale ratio parameter $\epsilon ~(0<\epsilon<1)$ in the Eqn. (\ref{eq.1}) as follows 
\begin{equation}
\begin{array}{l}\label{eq.6}	
\epsilon \dot{x} =-x+y+z,\\
\dot{y} =xy-z,\\
\dot{z} =-axz+y+b,\\
\end{array}
\end{equation}
so that $x$ is fast variable and $y, z$ are slow variables. We rescale the above Eqn. (\ref{eq.6}) by setting $\tau=t/\epsilon$ as 
\begin{equation}
\begin{array}{l}\label{eq.7}	
x^{'} =-x+y+z,\\
y^{'} =\epsilon(xy-z),\\
z^{'} =\epsilon(-axz+y+b),\\
\end{array}
\end{equation}
where ${x^{'}}\equiv\frac{dx}{d\tau}$. By setting $\epsilon=0$, the trajectory of Eqn. (\ref{eq.6}) converges to the solution of the differential algebraic equation  
\begin{equation}
\begin{array}{l}\label{eq.8}	
0 =-x+y+z,\\
\dot{y} =xy-z,\\
\dot{z} =-axz+y+b.\\
\end{array}
\end{equation}

\begin{figure}[ht]
	\centerline{
		\includegraphics[scale=0.55]{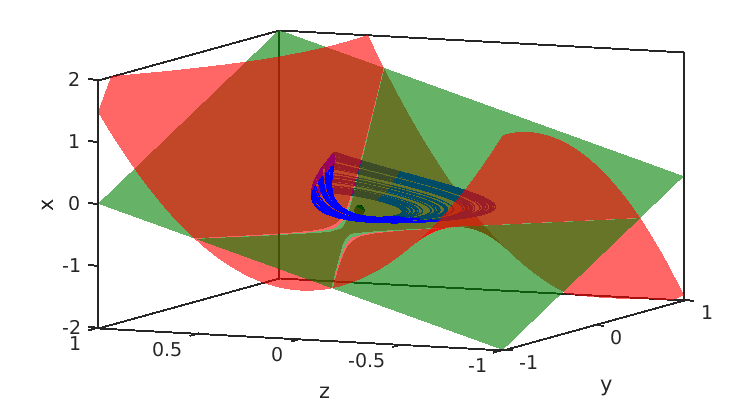} }
	\caption{Plot of  the critical manifold $S_0$ and slow manifold $S_{\epsilon}$ at $\epsilon=0.85$ are illustrated by green and red colors, respectively. Chaotic attractor (blue line) is drawn for $a=3.17$ and $b=0.26$. The position of the unstable fixed point (solid dot): $(-0.2420, -0.2286, 0.0553)$.}
	\label{fig6} 
\end{figure}

So,  as per above definition, $S_0=\{(x,y,z)\in \mathbb R^3 : x=y+z\}$ is called a critical manifold. From Fenichel theory \cite{fenichel1979}, we know that a subset $S_{\epsilon}$ of $S_{0}$ exists for $\epsilon~(>0)$ if $S_{\epsilon}$ is normally hyperbolic. Our system is normally hyperbolic because of real part of $\frac{\partial (-x+y+z)}{\partial x}=-1 (\ne 0), (x,y,z)\in S_{\epsilon}$. So there exists a local invariant manifold (slow manifold) $S_{\epsilon}$ and we derive this equation by using geometric singular perturbation method \cite{fenichel1979, guckenheimer2008, desroches2012}.   
\par In $S_0$, relation among variables $x,y$ and $z$ is $x=y+z$. Let in $S_{\epsilon}$, this relation becomes
\begin{equation}
\begin{array}{l}\label{eq.9}	
x=y+z+O(\epsilon).
\end{array}
\end{equation}
By Taylor series expansion in $\epsilon$, we get  
\begin{equation}
\begin{array}{l}\label{eq.10}	
x=y+z+\epsilon H(y,z)+O(\epsilon^2).
\end{array}
\end{equation}
Differentiate Eqn. (\ref{eq.9}) and then multiply both sides by $\epsilon$, we have,
\begin{equation}
\begin{array}{l}\label{eq.11}
\epsilon \dot{x}=\epsilon \dot{y}+\epsilon \dot{z}+O(\epsilon^2).
\end{array}
\end{equation}
Using Eqns. (\ref{eq.6}) and (\ref{eq.9}) in Eqn. (\ref{eq.11}), we get 
\begin{equation}
\begin{array}{l}\label{eq.12}
\epsilon \dot{x}=\epsilon\Big (y^2-az^2+(1-a)yz+y-z+b\Big )+O(\epsilon^2).
\end{array}
\end{equation}
Again from first equation of Eqn. (\ref{eq.6}), 
\[
\epsilon \dot{x} =-x+y+z.
\]
Now using Eqn. (\ref{eq.10}) in above equation, we get 
\begin{equation}
\begin{array}{l}\label{eq.13}
\epsilon \dot{x}=-(\epsilon H(y, z)+O(\epsilon^2)).
\end{array}
\end{equation}
Comparing Eqns. (\ref{eq.12}) and (\ref{eq.13}), finally we obtain\\
\begin{equation}
\begin{array}{l}\label{eq.14}
H(y,z)=-\Big (y^2-az^2+(1-a)yz+y-z+b\Big ).
\end{array}
\end{equation}
\begin{figure*}[ht]
	\centerline{\includegraphics[scale=0.05]{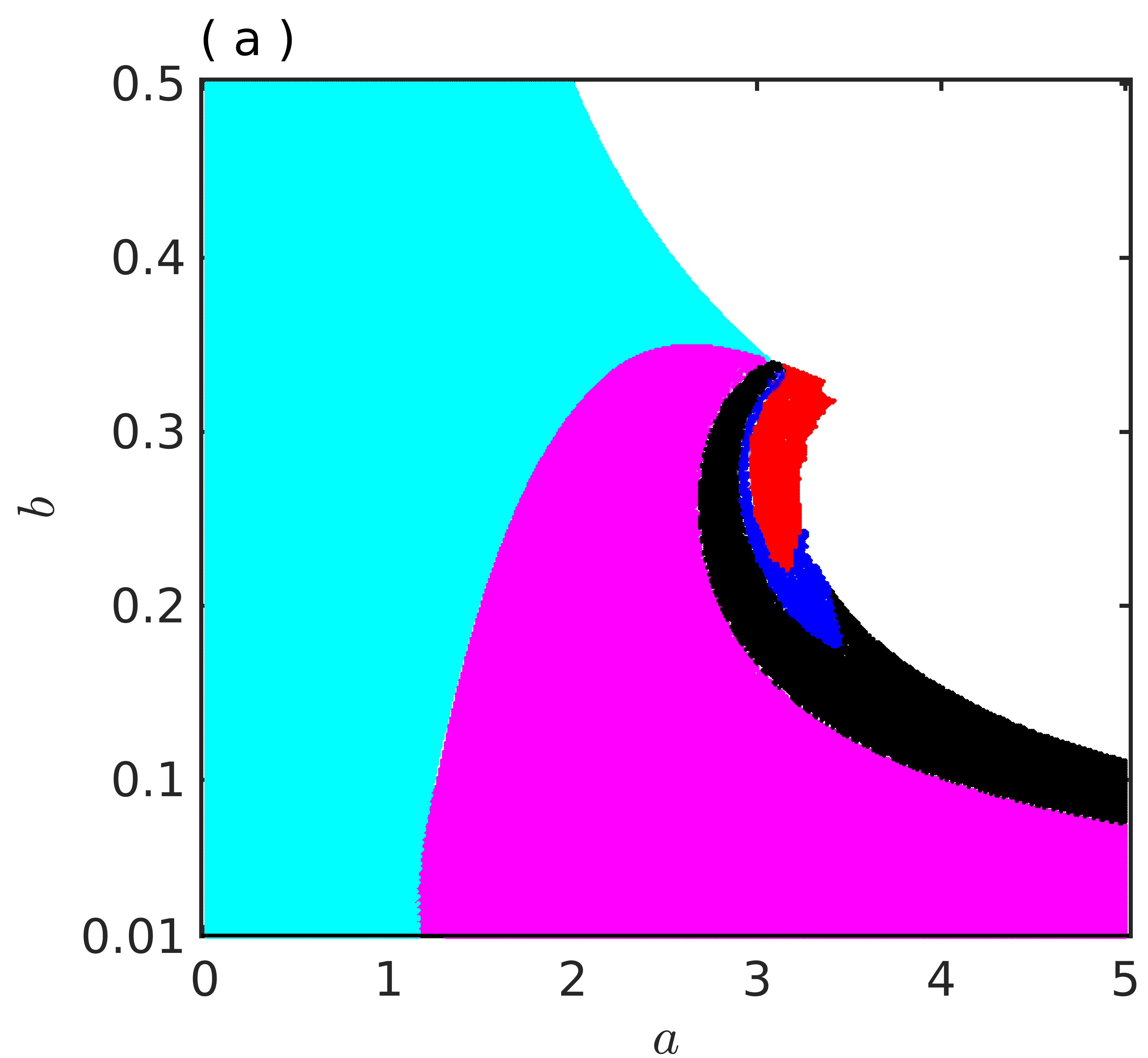}
		\includegraphics[scale=0.5]{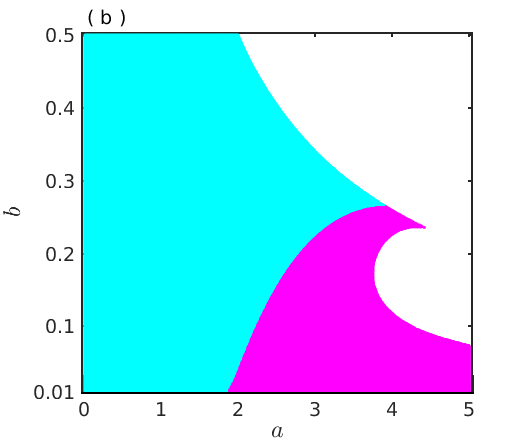}}
	\centerline{\includegraphics[scale=0.6]{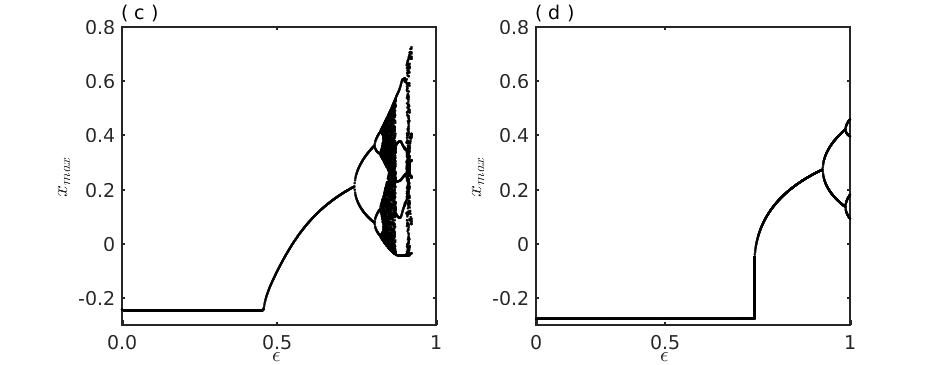}}
	\caption{Upper panel:  different dynamical states in $a$-$b$ parameter spaces for (a) $\epsilon=0.9$, and (b) $\epsilon=0.1$. Here, Cyan: stable steady state, magenta: period-1, black: period-2, blue: period-4, and red: greater than period-4 or chaotic state. White region is for unbounded solution. Lower panel: bifurcation diagrams with respect to the time-scale ratio parameter $\epsilon$ for (c) $\{a, b\}=\{3.17, 0.26\}$  and (d) $\{a, b\}=\{2.68, 0.304\}$. The initial condition is $x(0)=0.3, y(0)=0.2$, and $z(0)=0.1$.  From this figure, it is noticed that chaotic state is annihilated by decreasing the value of time-scale parameter $\epsilon$ and stable steady state and periodic solutions dominate the whole parameter region.}
	\label{fig7} 
\end{figure*}  
\par We use the above value of $H(y, z)$ in Eqn. (\ref{eq.10}) to calculate the approximated slow manifold $S_{\epsilon}$. We plot the critical manifold $S_0$ (in green color), an approximated slow manifold $S_{\epsilon}$ (in red color) in Fig.\ \ref{fig6}. The chaotic attractor for the set of values $a=3.17, b=0.26$ and $\epsilon=0.85$ is shown by blue line. This attractor lies on the both sides of the manifold $S_0$ and $S_{\epsilon}$. We take the value $\epsilon=0.85$ for $S_{\epsilon}$ (though this value is not so small enough, but we choose this value due to the appearance of chaos).     
\par Next we study the effect of time-scale parameter $\epsilon$ of the system (\ref{eq.6}) in ($a$-$b$) parameter space for the previously mentioned range. For this, we first set $\epsilon=0.9$ and different dynamical states like, stable fixed point, period-1, period-2, period-4 and greater than period-4 or chaos are shown in Fig.\ \ref{fig7}(a). The stable fixed point dominates in the ($a$-$b$) parameter space compare to Fig.\ \ref{fig5} for $\epsilon=1.0$ case, i.e., without any time-scale.   Also the region of chaotic state  decreases compare to the system (\ref{eq.1}) (in Fig.\ \ref{fig5}). If we decrease the value of time-scale parameter $\epsilon$, then the chaotic region is completely annihilated in the ($a$-$b$) parameter space. Figure \ref{fig7}(b) shows that the region of chaotic state is completely annihilated whereas region of stable fixed point is expanded and rest parts of the region are filled by period-1 limit cycle for $\epsilon=0.1$. So by introducing the slow-fast parameter $\epsilon$ in our original system (\ref{eq.1}), we can control chaos. Now we take two sets of parameter values of $\{a, b\}, \ i.e.,$ $\{3.17, 0.26\}$ and $\{2.68, 0.304\}$ from the Figs.\ \ref{fig7}(a, b), we have illustrated two bifurcation diagrams by varying slow-fast ratio parameter $\epsilon \in[0,1]$ for understanding the various change of dynamics in slow-fast system in Figs.\ \ref{fig7}(c, d), respectively. For the first set of parameter values,  the period-doubling route to chaos is observed, whereas for second case, the chaotic behavior in the system is terminated and two states (stable fixed point and period-1 limit cycle) are dominated.            
\section{Conclusions}
\par To summarize, in this paper, we have proposed a new simple chaotic system with two nonlinear terms. We have elaborately discussed the nature of fixed points through linear stability analysis. We have also analytically derived the first Lyapunov coefficient to show the nature of Hopf bifurcation by which fixed point changes its stability. Using bifurcation diagram, we have investigated the period-doubling route to chaos and also verified the results using Lyapunov exponent and first return map. We also separate the region of different dynamical behavior in the parameter space. Finally, we introduce slow-fastness in our model and try to understand the regarding changes of dynamics. Variation of dynamics has been captured in ($a$-$b$) parameter space for different values of the slow-fast parameter $\epsilon$. Most interesting fact is that we can control the chaotic and higher periodic orbits by adjusting the value of $\epsilon$.% that the chaotic region is fully controlled by interplaying two different time-scale in transformed slow-fast system.   
\\
\par {\bf Acknowledgments:}
The authors would like to thank Syamal K. Dana, Chittaranjan Hens, Gourab Kumar Sar, Sayeed Anwar, Soumen Majhi, Srilena Kundu, Sayantan Nag Chowdhury, Subrata Ghosh and Sarbendu Rakshit for helpful discussions and comments. 
\section*{Appendix: Calculation of first Lyapunov coefficient} 
\par Let us consider the differential equation 
\begin{equation}
\begin{array}{l}\label{eq.15}
\dot{x}=f(x,\alpha),
\end{array}
\end{equation}
where $x\in \mathbb{R}^3$ and $\alpha\in \mathbb{R}^1$ represent the state variables and system parameter respectively, and $f:\mathbb{R}^3\times \mathbb{R}^1 \rightarrow \mathbb{R}^1$ is a smooth function. Let the Eqn. (\ref{eq.15}) has a fixed point $x=x_0$ at $\alpha=\alpha_0$. Now, replacing $x-x_0$ by $X$ and Eqn. (\ref{eq.15}) becomes
\begin{equation}
\begin{array}{l}\label{eq.16}
\dot{X}=F(X)=f(X,\alpha_0).
\end{array}
\end{equation}
Here, $F(X)$ is also a smooth function and we can expand it in Taylor series in terms of symmetric functions as follows,
\begin{equation}
\begin{array}{l}\label{eq.17}
F(X)=AX+\frac{1}{2!}B(X,X)+\frac{1}{3!}C(X,X,X)+O(||X||^4),
\end{array}
\end{equation}     
where $A=f_x(0,\alpha_0)$ is the Jacobian matrix and for i=1,2,3, 
\begin{equation}
\begin{array}{l}\label{eq.18}
B_{i}(X,Y)=\sum_{j,k=1}^{3} \frac{\partial^2{F_i(\xi)}}{\partial \xi_{j} \partial\xi_{k}}|_{\xi=0} X_j X_k,\\
C_{i}(X,Y,Z)=\sum_{j,k,l=1}^{3} \frac{\partial^3{F_i(\xi)}}{\partial \xi_{j} \partial\xi_{k} \partial\xi_{l}}|_{\xi=0} X_j X_k X_l.
\end{array}
\end{equation} 
In our case, multilinear symmetric functions are\\
\begin{equation}
\begin{array}{l}\label{eq.19}
B(X,Y)=(0,x_1y_2+x_2y_1,-a^*(x_1y_3+x_3y_1)),\\
C(X,Y,Z)=(0,0,0).
\end{array}
\end{equation}

\par We assume that A has a pair of complex eigenvalues  $\lambda_{2,3}=\pm iw_0(w_0>0)$ on the imaginary axis at $(x_0,\alpha_0)$. Let generalized eigen space of $A$ be $T^C$ which is the largest invariant subspace spanned by eigenvectors corresponding to $\lambda_{2,3}$. Let $p,q\in$ C$^3$ be such eigen vectors then  
\begin{equation}
\begin{array}{l}\label{eq.20}
Aq=iw_0q,\\
A^Tp=iw_0p,\\
<p,q>=\sum_{i=1}^{3}\bar{p}_iq_i=1,
\end{array}
\end{equation}
where $A^T$ denotes the transpose of the matrix $A$ and $<...>$ is the standard scalar product over $\mathbb{C}^3$. For our system (\ref{eq.1}), we obtain\\
$
q\approx(-0.0404 -0.5130i,~ 0.6402,~ -0.1244 - 0.5568i),\\
p \approx(-0.1629 - 0.0591i,~ 0.78 - 0.1677i,~ 0.1501 - 0.8436i),\\
$
where $w_0\approx 1.0843$.
\par  Now we choose any $y\in T^C$ which is represented as $y=wq+\bar w\bar q$, where $w=<p,q>\in \mathbb{C}$. The two-dimensional center manifold associated with the eigenvalues $\lambda_{2,3}= \pm iw_{0}$ can be parameterized by $w$ and $\bar w$ in the form $X=H(w,\bar w)$. Here $H: \mathbb{C}^2\rightarrow\mathbb{R}^3$ can be expanded as follows
\begin{equation}
\begin{array}{l}\label{eq.21}
H(w,\bar w)=wq+\bar w\bar q	+\sum_{2\le j+k\le 3}\frac{1}{j!k!}h_{jk}w^j\bar w^k+O(|w|^4),
\end{array}
\end{equation}
where $h_{jk}\in \mathbb{C}^3$ and $h_{jk}=\bar h_{kj}$. Differentiating $H(w,\bar w)$ with respect to $t$ and substituting this into Eqn. (\ref{eq.15}) and using Eqn. (\ref{eq.16}), we get the following differential equation
\begin{equation}
\begin{array}{l}\label{eq.22}
H_{w}\dot{w}+H_{\bar w}\dot{\bar w}=F(H(w,\bar w)).
\end{array}
\end{equation}  
Using Eqns. (\ref{eq.16}), (\ref{eq.21}) and (\ref{eq.22}), we determine the complex coefficients $h_{ij}$ and from the Eqn. (\ref{eq.22}), the $w$  evolves on the center manifold,\\
\begin{equation}
\begin{array}{l}\label{eq.23}
\dot{w}=iw_0w+\frac{1}{2}G_{21}w|w|^2+O(|w|^4),
\end{array}
\end{equation}
where $G_{21}\in \mathbb{C}$.
Substituting Eqn. (\ref{eq.23}) into Eqn. (\ref{eq.22}) and using Eqns.\~ (\ref{eq.17}) and (\ref{eq.18}), we get the following coefficients,
\begin{equation}
\begin{array}{l}\label{eq.24}
h_{11}=-A^{-1}B(q,\bar q),\\
h_{20}=(2iw_0I_3-A)^{-1}B(q,q),
\end{array}
\end{equation}
where $I_3$ is the 3 $\times$ 3 identity matrix. Finally $G_{21}$ is obtained from the condition of existence of solution of the equation for $h_{21}$. So, 
\begin{equation*}
	\begin{array}{l}
		G_{21}=<p,~C(q,q,\bar q)+B(\bar q,h_{20})+2B(q,h_{11})>,
	\end{array}
\end{equation*}
where we use $<p,q>=1$. So the first Lyapunov coefficient is defined as 
\[
l_1=\frac{1}{2}Re(G_{21}),
\]
which determines the nonlinear stability of a nondegenerate codimension one Hopf bifurcation.
Now, for calculating $l_1$, we first find  \\
$ h11\approx(   0.6348, 1.0854, -0.4507),$\\
$h20\approx( -0.2273 + 0.0747i, -0.1643 - 0.1240i, -0.2251 - 0.2942i).$
using Eqns. (\ref{eq.18}), (\ref{eq.24}) and (\ref{eq.24}).
Finally, we get $l_1\approx0.04196~(> 0)$.

\end{document}